# Effect of FABr Over-Stoichiometry on the Morphology and Optoelectronic Properties of Wide-Bandgap $FAPbBr_3$ Films

G. Ammirati[1], F. Martelli[1], F. Toschi[1], S. Turchini[1], P. O'Keeffe[2], A. Paladini[2], F. Matteocci[3], J. Barichello[1], D. Di Girolamo[3], S. Piccirillo[4], A. Di Carlo[1,3] and D. Catone[1,*].

*[1]Istituto di Struttura della Materia - CNR (ISM-CNR), EuroFEL Support Laboratory (EFSL), Via del Fosso del Cavaliere 100, 00133, Rome, Italy.*
*[2]Istituto di Struttura della Materia - CNR (ISM-CNR), EuroFEL Support Laboratory (EFSL), Monterotondo Scalo 00015, Italy.*
*[3]Centre for Hybrid and Organic Solar Energy (CHOSE), Department of Electronic Engineering, Tor Vergata University of Rome, Via del Politecnico 1, 00133, Rome, Italy.*
*[4]Department of Chemical Sciences and Technologies, Tor Vergata University of Rome Via della Ricerca Scientifica, 00133, Rome, Italy.*

*Corresponding author: daniele.catone@cnr.it.

## Abstract

In this study, we investigate the impact of formamidinium bromide (FABr) over-stoichiometry in the precursor solution on the optoelectronic properties and morphology of the resulting films of formamidinium lead bromide ($FAPbBr_3$). Optical characterization, including steady-state absorption, photoluminescence (PL), and femtosecond transient absorption spectroscopy, reveals a systematic blueshift in emission energy with increasing FABr content, attributed to the passivation of bromine vacancies and to the reduction of defect-assisted recombination. Power-dependent PL confirms this interpretation: the stoichiometric film exhibits a PL band due to donor–acceptor pair (DAP) recombination as identified by the typical excitation-dependent blueshift, whereas FABr-enriched samples show no evidence of DAP emission, indicating effective defect passivation. Additionally, morphological characterization shows a reduction in grain size with increasing FABr excess, indicating a trade-off between improved electronic quality and enhanced structural disorder. The film synthesized with a 5% excess of FABr provides the optimal balance, yielding the highest power conversion efficiency (6.26%), average visible transmittance (61.6%), and light utilization efficiency (3.85%). These results demonstrate that fine-tuning the precursor stoichiometry through controlled FABr addition represents a simple yet effective strategy to enhance the optoelectronic quality and performance of semitransparent perovskite solar cells.

## Introduction

Hybrid metal halide perovskites (MHPs) have revolutionized photovoltaic technology over the last decade, primarily due to their exceptional optoelectronic properties. MHPs show promising physical properties for photovoltaics, such as high absorption coefficients in the visible range (~$10^5$ $cm^{-1}$), band gap tunability, long carrier diffusion length ( >1 μm), long carrier lifetime ( ~ 1 μs), and unusually high defect tolerance.[1–6] These properties counterbalance the moderate carrier mobility (~10 $cm^2$ $V^{-1}$ $s^{-1}$), which is lower with respect to traditional semiconductors (e.g., 1000 $cm^2$ $V^{-1}$ $s^{-1}$ for Si).[7] MHPs for photovoltaic applications are usually employed as a thin film of a few hundred nanometers and, in these cases, take the form of a polycrystalline structure composed of micrometer-sized grains. These features collectively make halide perovskites highly attractive for a variety of optoelectronic applications, including photovoltaics, LEDs and photodetectors. Furthermore, the films are usually obtained by means of low-cost fabrication methods such as spin coating, blade coating, screen printing, or inkjet printing.[8–11] Consequently, an extensive research effort has been dedicated to optimizing their compositional engineering, morphology control, and device integration.

Among the family of perovskite absorbers, wide-bandgap compositions such as formamidinium lead bromide ($FAPbBr_3$) have gained particular interest, especially for applications in semitransparent solar cells and tandem solar cell configurations. The large bandgap (~2.3 eV) of $FAPbBr_3$ is particularly well suited to harvesting the high-energy portion of the solar spectrum, making this material an ideal candidate as a top-cell absorber in tandem photovoltaic devices.[12,13] Nonetheless, a significant challenge in photovoltaic applications of wide-bandgap perovskites is the substantial reduction in open-circuit voltage compared to their optical bandgap.[14] This voltage deficit primarily originates from unfavorable alignment of energy levels between the absorber and the adjacent charge transport layers, as well as significant non-radiative recombination within the active material.[15] In particular, non-radiative recombination is associated with grain boundaries, halide vacancies, interstitial defects, and other point defects.[16,17] Consequently, one of the main challenges in achieving high efficiency and stability of $FAPbBr_3$-based photovoltaic devices lies in the careful management of the interplay between crystallinity, film morphology, defect passivation, and energy-level alignment to minimize such performance losses.

In this context, precursor engineering via controlled over-stoichiometry has emerged as an effective strategy to improve the optoelectronic quality of metal–halide perovskites. For instance, $MAPbI_3$ films fabricated with an excess of methylammonium iodide have been shown to form defect-passivating surface layers, leading to suppression of non-radiative recombination and enhancement of charge mobility.[18] Similarly, in $FAPbBr_3$ nanocrystals, the introduction of excess formamidinium bromide has been reported to refill bromine and formamidinium vacancies, resulting in a substantial increase in photoluminescence quantum yield and more balanced charge transport due to a reduced density of insulating organic ligands.[19] Despite these promising results in $MAPbI_3$ thin films and $FAPbBr_3$ nanocrystals, the role of precursor over-stoichiometry in governing the polycrystalline morphology, defect landscape, and electronic properties of $FAPbBr_3$ thin films specifically tailored for photovoltaic applications remains insufficiently explored.

In this paper, we investigate how different levels of FABr excess (+5% and +10%) in the synthesis of the active material affect the crystallization and the optical properties of the $FAPbBr_3$ thin films. By combining morphological characterization, steady-state and temperature-dependent photoluminescence (PL), optical absorption studies, and transient absorption measurements, we aim to provide deeper insights into the role of the FABr stoichiometry employed during the synthesis of optimized wide-bandgap perovskites for semitransparent photovoltaic applications.

## Experimental Methods

### Sample preparation, synthesis of the materials and device fabrication

$FAPbBr_3$ thin films were deposited onto ITO conductive substrates (Kintec company, 1.1 mm thick, 10 Ω/sq of sheet resistance). The substrates were first cut into squares of 2.5 x 2.5 cm and then washed with a sequential cleaning procedure first in a soap (5% in volume) and water solution, then in deionized water, and finally in isopropanol through ultrasonic baths of 10 minutes to remove organic compounds and dust. Before depositing the perovskite, the cleaned substrates were soaked under a UV-light treatment for 30 minutes to improve the surface wettability to obtain a good perovskite film deposition. In a vial, 1.3 M FABr (Dyesol) and 1.3 M $PbBr_2$ (TCI) were dissolved in DMSO and were left stirring all night. $FAPbBr_3$ deposition was performed under an $N_2$ atmosphere. The substrates were first heated at 60°C, and then, 80 μL of the perovskite solution was dropped on the surface and spin-coated at 4000 rpm for 20 seconds. The solvent quenching method was employed to enhance the perovskite crystallization, and 200 μL of ethyl acetate was dropped 10 seconds after the initiation of the spin. The substrates were then annealed at 80 °C for 10 minutes. This procedure was followed to fabricate the stochiometric sample of $FAPbBr_3$ (Control) while for the samples synthesized with an excess of formamidine bromide (FABr%), concentrations of 1.365 M FABr (a 5% excess) and 1.43 M FABr (a 10% excess) were used for the growth of the FABr 5% and FABr 10% samples, respectively.

The solar cell devices were fabricated using perovskite films prepared as previously described. The remaining layers of the photovoltaic devices—namely, the electron transport layer, the hole transport layer, and the top electrode—were deposited following the procedures reported in [20].

## Steady-state optical spectroscopy

The absorbance spectra of thin films were acquired with a Shimadzu UV-VIS spectrometer (model UV-2700) in combination with the Shimadzu Multipurpose Large-Sample Compartment (MPC-2600A) in the 220-850 nm range. These units were used for transmittance and diffuse and specular reflectance measurements.

The average visible transmittance (AVT) of the semitransparent devices was determined following the procedure outlined in ISO 9050:2003 and using. See details in **Section S.1** of Supporting Information (SI).

## SEM and XRD

Field emission scanning electron microscopy (FESEM - ZEISS SIGMA 300) was used for the morphological characterization of the samples, operating at an acceleration voltage of 10 kV and working distance of 13 mm. The samples were mounted on aluminum stubs using conductive carbon tape and the images were acquired at the same magnification (25kX), with the aim of comparing their grain-size distributions, which were analyzed using Gwyddion software.[21]

X-Ray diffraction (XRD) patterns were acquired using a Rigaku SmartLab SE diffractometer equipped with a Cu K$\alpha$ radiation source, in the 2θ range of 5–60°.

## Photoluminescence

PL was performed at 12K with the help of a closed-cycle He cryostat and a PID temperature controller was used to regulate at all intermediate temperatures between RT and 12K. The samples were excited using a continuous-wave laser at 405 nm supplied by the Matchbox 2 laser series. The laser beam is focused on the sample with a spot size of 100 μm, and the PL is collected by a telescope into the monochromator (HORIBA Jobin Yvon – iHR320) with a focal length of 320 mm, and a spectral range of 350 nm - 1100 nm with a 1200 gr/mm grating. Suitable filters were used to eliminate the laser light scattered by the sample.

## Femtosecond Transient Absorption Spectroscopy

Femtosecond transient absorption spectroscopy (FTAS) was carried out using an amplified femtosecond laser source delivering 35 fs pulses at a 1 kHz repetition rate with an average power of 4 W, centered at 800 nm. Tunable pump pulses were produced via an optical parametric amplifier. A white-light supercontinuum probe covering the 360–780 nm range was obtained by focusing approximately 3 μJ of the 800 nm beam into a rotating $CaF_2$ crystal. Both the pump and probe beams were focused onto the sample, with spot diameters of about 200 μm and 150 μm, respectively. The time delay between pump and probe pulses was adjusted by varying the probe optical path length, yielding an overall instrument response of roughly 50 fs. Further experimental details are available elsewhere.[22,23]

## Results and Discussion

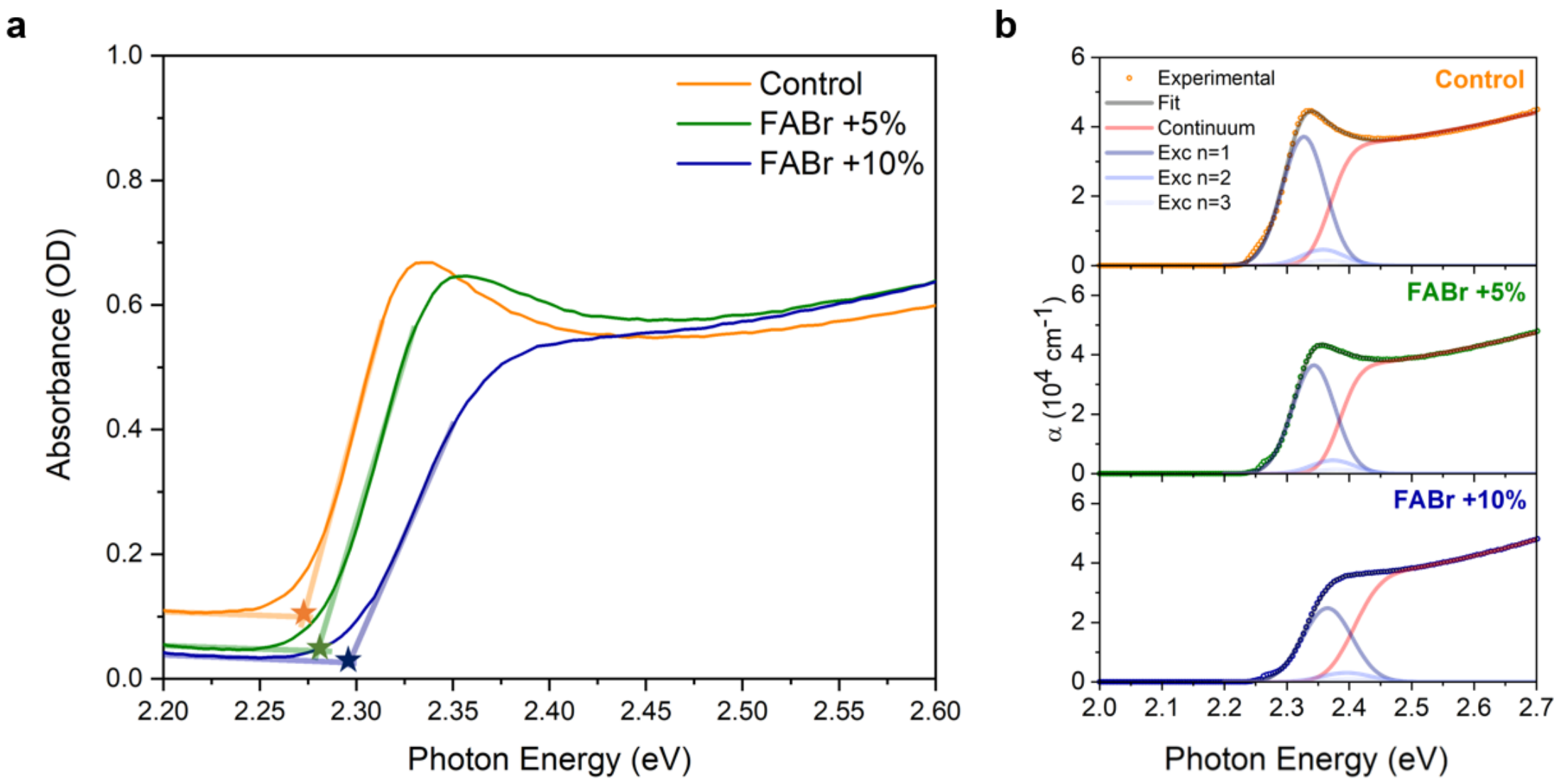

**Figure 1:** (a) Steady-state absorbance spectra of the Control, FABr +5% and FABr +10% samples (see text for details); (b) absorption coefficient of the Control, FABr +5%, and FABr +10% films (orange, green and blue scatter, respectively). The data are fitted using the Elliott model (black line), which is further decomposed into contributions from the continuum states (red line) and the three excitonic states (blue line). See text for the details.

**Figure 1a** shows the absorbance for the Control ($FAPbBr_3$ stoichiometrically synthesized), FABr +5% ($FAPbBr_3$ synthesized with a 5% excess of FABr), and FABr +10% ($FAPbBr_3$ synthesized with a 10% excess of FABr) films, showing a blueshift of the optical onset of 2.273 eV, 2.280 eV, and 2.296 eV with the increase of the FABr excess. The onset of the signal was set at the intercept (orange, green, and blue stars respectively for Control, FABr +5%, and FABr +10%) between the absorption peak and the low-energy tail related to the background scattering. The steady-state absorption spectrum of the Control sample exhibits a distinct excitonic feature, but as the FABr excess increases, this feature becomes less pronounced. **Figure 1b** reports the corresponding absorption coefficient for the studied samples. The absorption coefficient for each thin film was estimated on the basis of the transmittivity (T), reflectivity (R), and thickness of the samples ($t$) according to **Equation 1**:

$$\alpha(\hbar\omega) = \frac{-\ln\left(\frac{T(\%)}{100 - R(\%)}\right)}{t} \tag{1}$$

The thickness was estimated to be approximately 150 nm.[20] The sub-bandgap intensity is set to zero to remove the scattering contribution to the spectra. Each steady-state absorption spectrum was decomposed into two components: an excitonic transition and a continuum electronic transition, both convolved with Gaussian broadening functions. Details of the fitting procedure following an Elliot-based excitonic absorption model are reported in **Section S.2** of SI and in our previous works.[24,25]

From this analysis, several key optical parameters were extracted, including the exciton binding energy ($E_b$), the electronic bandgap ($E_g$), Gaussian broadening of the excitonic ($\sigma_x$) and continuum ($\sigma_c$) contributions, as well as the coefficient $b$ that takes into account the non-parabolicity of the bands. In the fitting procedure, the exciton binding energy was kept fixed within 40 ± 5 meV, as variations within this range produced essentially identical results. All other parameters were left free to vary. The resulting values are summarized in **Table 1.**

| **Material** | **$E_g$ (eV)** | **$E_b$ (meV)** | **$\sigma_X$ (meV)** | **$\sigma_c$ (meV)** | **b($eV^{-1}$)** |
|---|---|---|---|---|---|
| Control | 2.368 | 40±5 | 34 | 26 | 0.69 |
| FABr +5% | 2.385 | 40±5 | 33 | 25 | 0.79 |
| FABr +10% | 2.406 | 40±5 | 39 | 33 | 0.91 |

**Table 1:** Absorption curves fitting parameters evaluated with the Elliott model as electronic bandgap ($E_g$), exciton binding energy ($E_b$), Gaussian broadening of the excitonic ($\sigma_x$ ) and continuum ($\sigma_c$ ) contributions, and the coefficient $b$.

The analysis of the absorption spectra reveals a clear evolution of the optical properties with increasing FABr incorporation. The electronic bandgap increases from 2.368 eV in the Control sample to 2.406 eV in the FABr +10% film, indicating a blue-shift of approximately 40 meV. This shift suggests that FABr incorporation modifies the electronic structure, likely through compositional or structural changes within the perovskite lattice. Since the fitting procedure yields comparable results within the range of 40 ± 5 meV, the exciton binding energy can be considered approximately constant in all samples, suggesting that the dielectric screening and effective masses governing excitonic stability are not strongly affected by the excess of FABr. More pronounced changes are instead observed in the broadening parameters: while the Control and FABr +5% films exhibit similar excitonic and continuum broadenings, both contributions increase significantly in the FABr +10% sample, as expected from the increase of compositional disorder. The increase of the coefficient *b* from 0.69 to 0.91 $eV^{-1}$ is consistent with the changes of the electronic structure induced by the FABr excess. Overall, FABr incorporation leads to a systematic bandgap widening while leaving the exciton binding energy almost unchanged.

Further investigation of the optical properties was performed using temperature- and power-dependent PL using a CW laser at 405 nm and at an irradiance of 100 $mW/cm^2$ unless otherwise specified. **Figure 2a** shows the PL spectra obtained at 12 K for the Control, FABr +5%, and FABr +10% thin films. The low-temperature measurements were performed to minimize thermal broadening of the PL line shape, thereby providing clearer insights into the defectivity of the samples. All spectra exhibit a main emission peak that undergoes a progressive blueshift with increasing FABr excess, consistent with the trend observed in the absorption spectra. Notably, the Control sample shows an additional broad emission band at lower energy, indicative of significant defect-related sub-bandgap states.

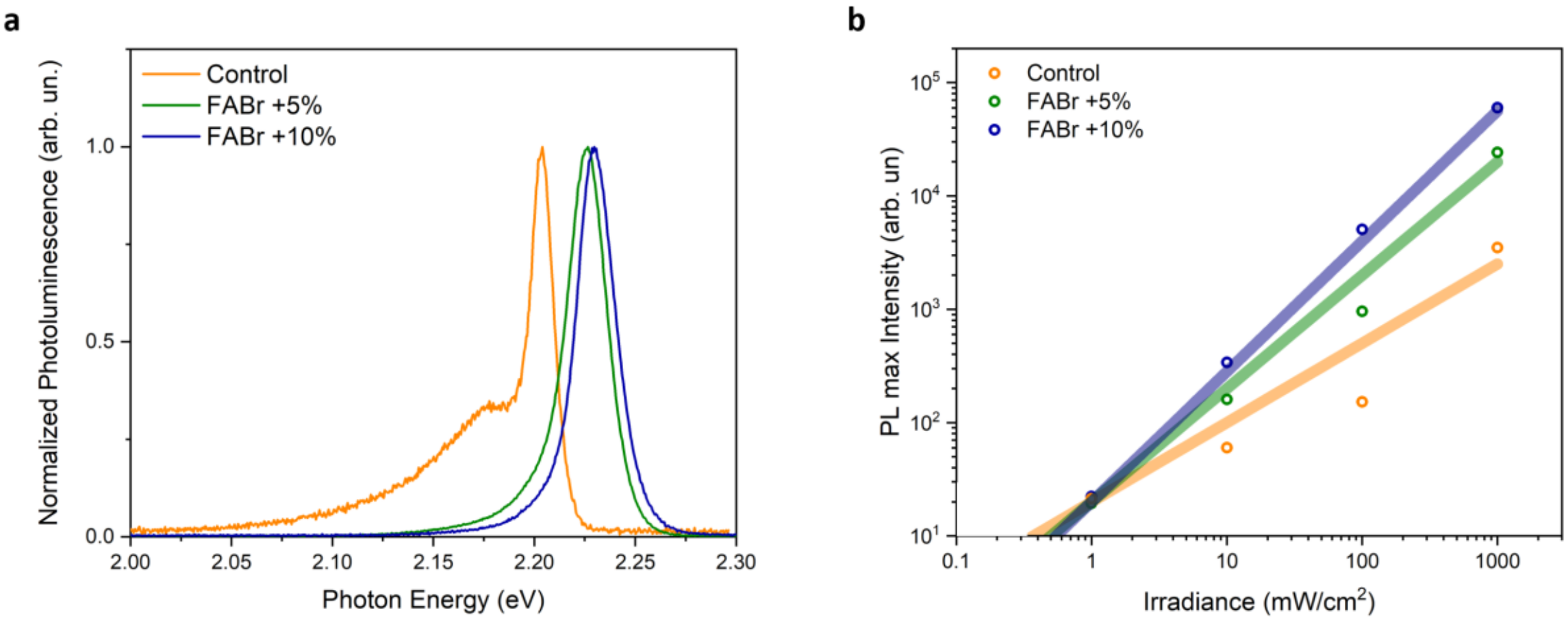

**Figure 2:** (a) PL spectra, normalized at the maximum intensity, acquired at 12 K for the Control, FABr +5% and FABr +10% thin films and with an excitation laser at 405 nm at an irradiance of 100 $mW/cm^2$; (b) PL main peak intensity as a function of the excitation laser irradiance for all samples.

Power-dependent PL measurements were performed to identify the recombination mechanisms responsible for the low-temperature emission, distinguishing among defect-mediated, excitonic, and free-carrier recombination. The dependence of the PL intensity ($I_{PL}$) on the excitation power ($I$) follows a power-law behavior, which in logarithmic form can be expressed as:

$$\log_{10}(I_{PL}) = \log_{10}(A) + n \log_{10}(I) \quad (2)$$

where $A$ is a proportionality constant and $n$ is the power-law exponent that provides an indication of the recombination mechanism involved.[26,27] Specifically, values of $n$<1 are typically associated with defect-mediated recombination processes, such as free-to-bound or donor-acceptor pair transitions. Conversely, $n$ close to 1 corresponds to purely excitonic recombination. When $n$ reaches the value of 2, the recombination process predominantly involves free carriers. Intermediate values between 1 and 2 indicate the coexistence of excitonic and free-carrier recombination within the sample. In light of these considerations, the extracted values of $n$ from power-dependent PL data provide direct insight into the nature of the recombination mechanisms governing the investigated materials. **Figure 2b** reports the PL intensity of the main emission peak as a function of the excitation irradiance for Control, FABr +5%, and FABr +10%, together with the best-fit curves obtained using **Equation 2**. The corresponding $n$ values are summarized in **Table 2.**

**Table 2:** Power-law exponent ($n$) obtained by fitting the PL intensity of the main emission peak as a function of the excitation irradiance for Control, FABr +5%, and FABr +10%.

| | **Control** | **FABr +5%** | **FABr +10%** |
|---|---|---|---|
| $n$ | 0.70±0.18 | 1.00±0.09 | 1.14±0.02 |

The numbers reported in Table 2 reveal changes of the dominant recombination mechanism upon FABr incorporation. The Control sample is primarily governed by defect-assisted recombination ($n$<1), while the FABr +5% and FABr +10% films exhibit power-law exponents ($n$≈1) consistent with predominant excitonic emission. The reduction of defect-assisted recombination in FABr +5% and FABr +10% films is consistent with FABr-induced passivation of bromine vacancies and in agreement with what is observed in absorption. This confirms that both FABr-enriched films maintain a predominantly excitonic character, suggesting that the variations observed in their absorption spectra principally arise from spectral broadening due to compositional disorder and thermal factors. FTAS and temperature-dependent PL measurements, reported in **Section S.3** and **Section S.4** of SI, further corroborate this interpretation.[28,29]

To further clarify the role of defect states in the PL recombination of the Control sample, **Figure 3a** displays the PL spectra recorded at 12 K and at selected excitations (0.001, 0.01, 0.1, 1, 10, 100, 1000, and 5000 mW cm$^{-2}$). Two distinct emission peaks are observed. The high-energy peak, centered at 2.20 eV, is attributed to free-carrier recombination and exhibits no noticeable shift with excitation power. In contrast, the low-energy emission shows a clear blueshift as the excitation irradiance increases. **Figure 3b** reports the evolution of the low-energy peak position (center of mass) as a function of excitation irradiance. This behavior is characteristic of donor–acceptor pair (DAP) recombination, involving polycentric bound-exciton complexes associated with two spatially close defect centers, where partial radiative recombination occurs. The energy of the emitted photons is expressed by the following equation:

$$E_{DAP} = (E_g - E_{D^0} - E_{A^0}) + \frac{e^2}{4\pi\epsilon_0\epsilon_r} \cdot \frac{1}{r_{DA}} \quad (3)$$

where $E_g$ is the bandgap, $E_{D0}$ and $E_{A0}$ are the binding energies of electrons and holes to their respective donor and acceptor centers, and the last term accounts for the Coulomb interaction between the ionized centers after recombination.[30–32]

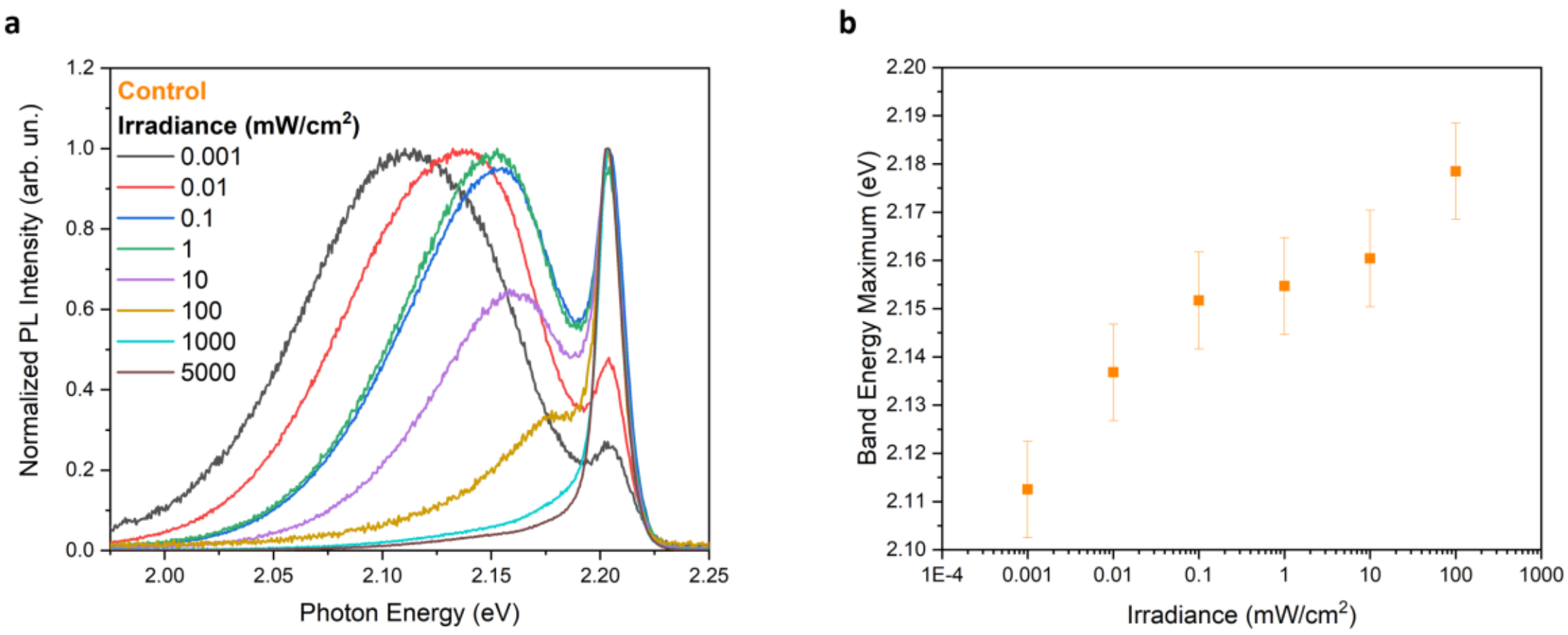

**Figure 3:** (a) PL spectra acquired at 12 K and normalized to their absolute maximum to facilitate comparison among different excitations; (b) band energy maximum of the low-energy PL peak of the Control sample obtained as a function of selected irradiances of excitation.

While the first three terms are independent of excitation power, the Coulombic term is power dependent. As the laser intensity increases, a larger number of donors and acceptors become occupied, leading to a reduction of their average distance $r_{DA}$. Consequently, the emission maximum of the DAP band exhibits a blueshift with increasing excitation power. This behavior is a fingerprint of DAP recombination and clearly different from excitation-induced thermal effects or other emission processes. The absence of a DAP band in the FABr +5% and FABr +10% samples indicates that FABr incorporation effectively reduces or passivates donors and acceptors.

These results indicate that the addition of FABr during film synthesis mitigates the defectivity of the films while promoting compositional and structural disorder. Since these two effects compete in determining the overall film quality, SEM characterization was performed to evaluate how FABr excess influences the morphology of the perovskite layers.

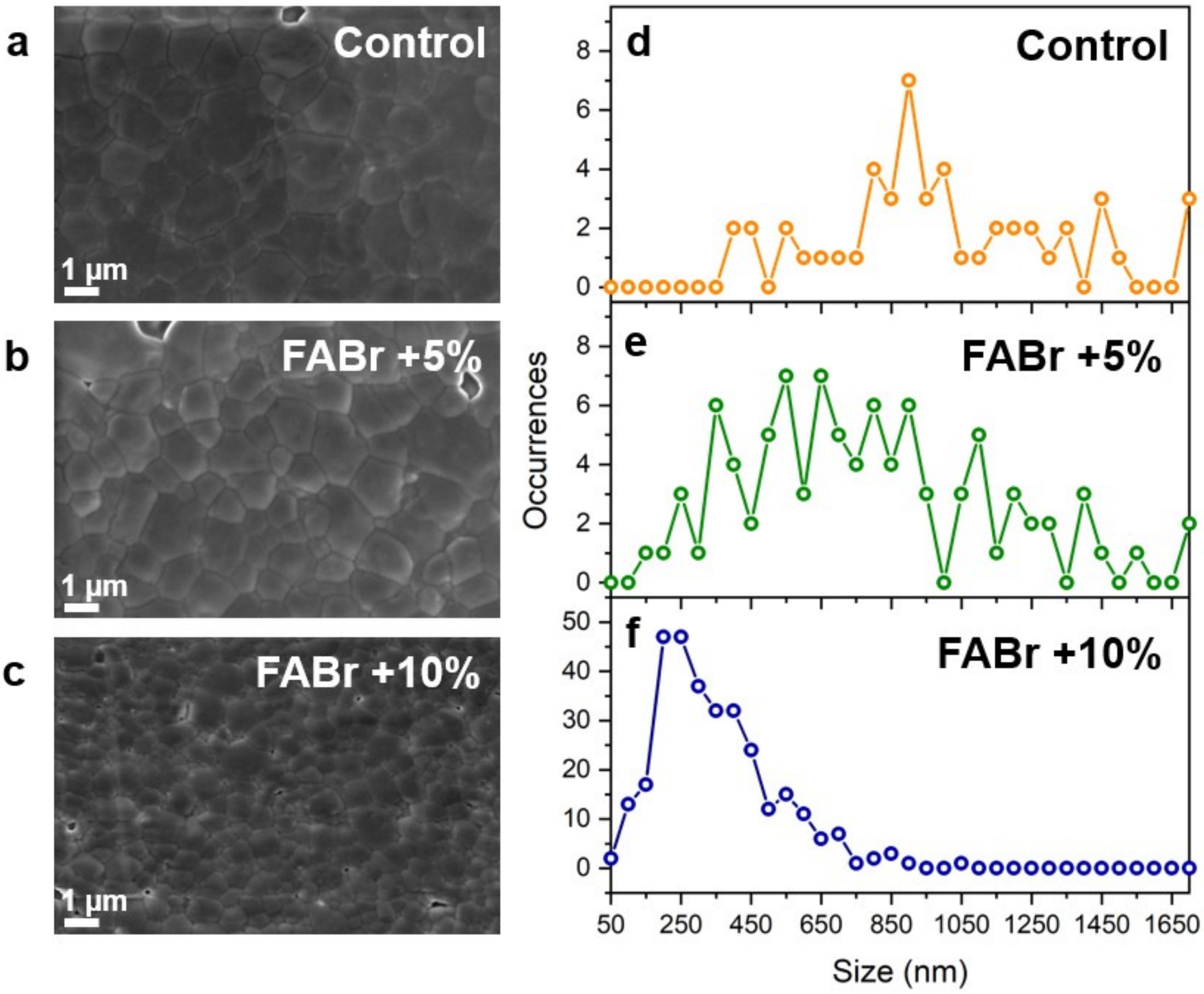

**Figure 4**: SEM images obtained for (a) Control, (b) FABr +5% and (c) FABr +10%. Analysis of the crystal sizes of (d) Control, (e) FABr +5% and (f) FABr +10%.

**Figure 4a-c** shows the top-view SEM images of Control, FABr +5% and FABr +10%. These images were analyzed using Gwyddion[21] to quantify the crystal size distribution in the studied samples, as reported in **Figure 4d-f**. Given the almost hexagonal shape of the grains, their size was estimated by measuring the longest distance between two opposite edges. The analysis of the SEM images reveals that the introduction of FABr excess significantly affects the crystallization of the material, resulting in a marked reduction of crystal size. The Control sample exhibits the largest grains, with an average size of approximately 850 ± 50 nm, while FABr +10% shows much smaller grains, around 250 ± 50 nm. The FABr +5%

sample exhibits a grain size distribution centered around 650 nm, spanning from 50 to 1200 nm, thus representing an intermediate morphology between the other two compositions. Overall, the excess of FA reduces the crystal and grain size distribution compared to the Control sample. While the SEM images highlight pronounced changes in grain size and morphology upon FABr addition, XRD measurements were employed to probe whether these variations are reflected in the average crystal structure of the films. The XRD patterns of Control, FABr +5% and FABr +10%, reported in **Figure S3** of SI, exhibit the same set of crystalline reflections that demonstrate that all the samples show an $\alpha$-$FAPbBr_3$ perovskite phase. No additional reflections or significant shifts in the peak positions are observed with the excess of FABr, indicating that the bulk crystal structure and the average lattice parameter of $FAPbBr_3$ are preserved.[20,33] Furthermore, the morphological changes revealed by the SEM analysis are in good agreement with the trends observed in the temperature-dependent PL (see **Section S.4** of SI). In particular, the broader FWHM of the PL emission in the FABr +10% sample correlates well with the reduced grain size and larger structural disorder visible in the SEM images. In addition, the PL peak energy of FABr +10% is consistently blue shifted relative to the Control at all measured temperatures, indicating that FABr excess alters the perovskite chemical composition in a way that significantly affects its optical properties.

To verify whether the spectroscopic findings are consistent with the performance of the photovoltaic devices, semitransparent solar cells based on the Control, FABr +5%, and FABr +10% films were fabricated. **Figure 5a-c** presents the current–voltage (J–V) curves, the external quantum efficiency (EQE) spectra and the transmittance for the three devices. See details in **Section S.6** of SI. The extracted photovoltaic parameters—short-circuit current density ($J_{SC}$), open-circuit voltage ($V_{OC}$), fill factor (FF), and power conversion efficiency (PCE)—together with the average visible transmittance (AVT) and light utilization efficiency (LUE), are summarized in **Table 3**. While PCE is the standard metric for evaluating photovoltaic performance, AVT and LUE are particularly important for semitransparent devices. AVT quantifies the average fraction of visible light transmitted through the device, whereas LUE, defined as the product of AVT and PCE, serves as a key figure of merit for semitransparent solar cells by capturing the trade-off between optical transparency and photovoltaic efficiency.

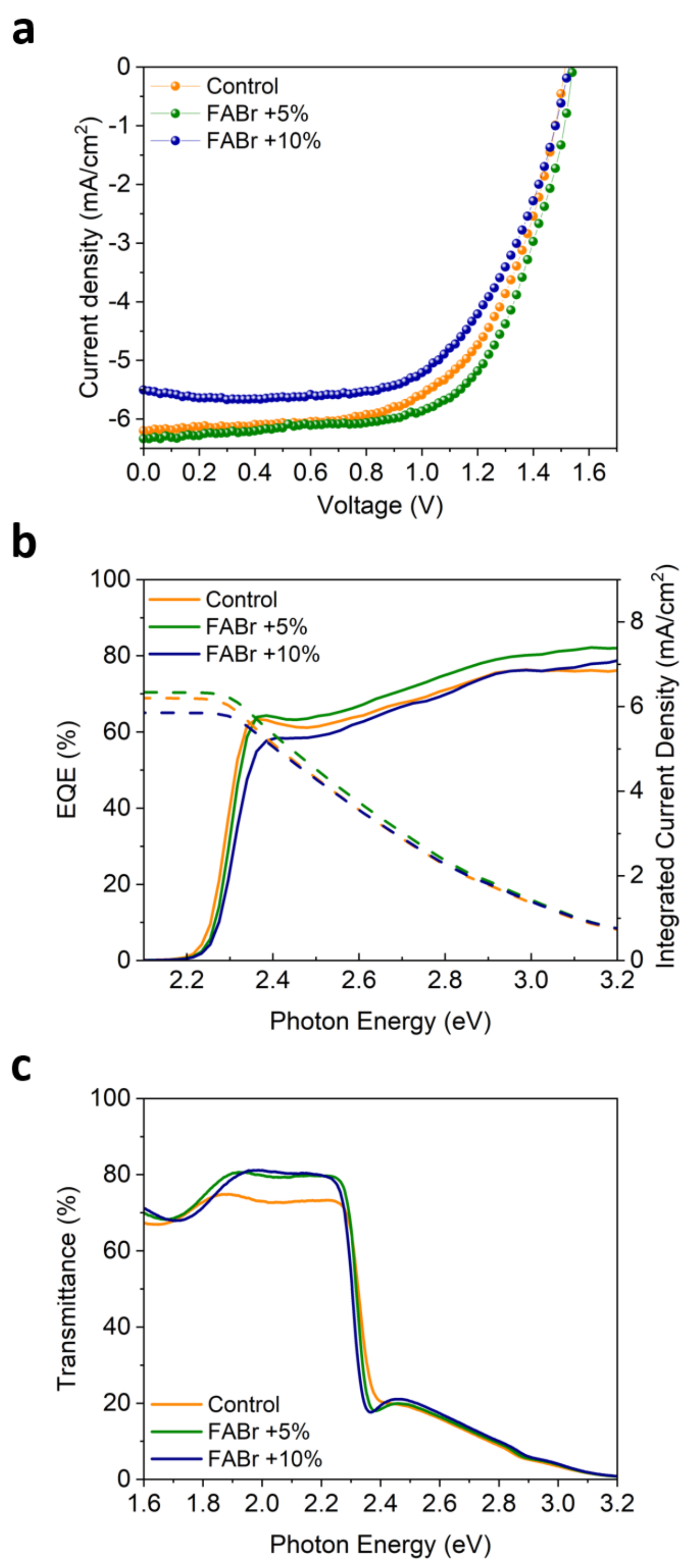

**Figure 5:** (a) J-V curve, (b) EQE (straight line) and Integrated Current Density (dashed line) and (c) transmittance obtained for Control, FABr +5%, and FABr +10%.

These results indicate that the FABr +5% sample achieves the optimal combination of PCE (6.26 %), AVT (61.6 %), and LUE (3.85 %), reflecting a favorable balance between two competing effects induced by FABr excess: effective defect passivation, increase of compositional disorder, and grain size reduction. This finding underscores the critical influence of precursor stoichiometry on film formation and its decisive impact on the photovoltaic performance of the resulting devices.

**Table 3**: Jsc, Voc, FF, PCE, AVT, and LUE measured for the solar cells fabricated using active layers synthesized with the same stoichiometry of Control, FABr +5%, and FABr +10%.

| **Material** | **Jsc (mA/cm²)** | **Voc (V)** | **FF (%)** | **PCE (%)** | **AVT (%)** | **LUE (%)** |
|---|---|---|---|---|---|---|
| Control | 6.20 | 1.51 | 61.52 | 5.77 | 60.2 | 3.47 |
| FABr +5% | 6.33 | 1.54 | 64.14 | 6.26 | 61.6 | 3.85 |
| FABr +10% | 5.50 | 1.53 | 62.94 | 5.29 | 58.4 | 3.09 |

## Conclusions

In this study, we explored the impact of different levels of FABr over-stoichiometry introduced during the fabrication of wide-bandgap $FAPbBr_3$ perovskite films on the resulting morphology, optoelectronic properties, and device performance. We have demonstrated that even small compositional variations significantly influence both the structural and optical behavior of the resulting films.

Spectroscopic analyses revealed a consistent blueshift in the emission energy and band gap upon FABr addition, primarily attributed to bromine vacancy passivation and the consequent suppression of defect-assisted recombination. In lead halide perovskites, bromine vacancies act as electron traps that locally distort the band structure, introducing sub-bandgap states and reducing the effective bandgap. The incorporation of FABr compensates for these vacancies, reducing the trap-state density and promoting a more uniform and defect-free lattice. As a result, non-radiative recombination is reduced—as evidenced by the absence of low-energy PL features and by the blue shift of the optical transition. The presence of donor–acceptor pair emission exclusively in the stoichiometric film further supports this interpretation. Its absence in FABr-enriched samples indicates effective defect passivation and the consequent suppression of sub-bandgap transitions.

The electron microscopy showed a clear reduction of grain size with increasing FABr content. The FABr 5% composition provided the best compromise between low defect density and controlled morphological disorder, resulting in the most favorable film quality. This is reflected in device performance, with FABr 5% delivering the highest power conversion efficiency (6.26 %), average visible transmittance (61.6 %), and light utilization efficiency (3.85 %).

In conclusion, we showed that careful precursor engineering, specifically the introduction of controlled amounts of precursor over-stoichiometry, can substantially influence crystallization, defect density, and thus the optical and electronic quality of perovskite films of $FAPbBr_3$. A slight excess of FABr represents a simple yet highly effective strategy to mitigate defect-related recombination while maintaining favorable film morphology, thereby enhancing the performance of semitransparent perovskite solar cells.

# Acknowledgements

S.T., G.A., D.C., P.O.K. acknowledges for funding the European Union – NextGenerationEU, M4C2, within the PNRR project NFFA-DI, CUP B53C22004310006, IR0000015. The authors acknowledge the European Project "Energy Harvesting in Cities with Transparent and Highly Efficient Window-Integrated Multi-Junction Solar Cells" (CITYSOLAR) for supporting the work, which received funding from the European Union's Horizon2020 research and innovation program under grant agreement number 101007084. J.B. acknowledge the support of the Project "Network 4 Energy Sustainable Transition—NEST", Spoke 1, Project code PE0000021, funded under the National Recovery and Resilience Plan (NRRP), Mission 4, Component 2, Investment 1.3—Call for tender No. 1561 of 11.10.2022 of Ministero dell'Università e della Ricerca (MUR) funded by the European Union—NextGenerationEU. The authors acknowledge Dr. Daniel Ory for sharing the fitting routine employed in the Elliot-based excitonic absorption model analysis of the absorption spectra.

# Author Contributions

G.A. and D.C. conceived the article. G.A., D.C., P.O.K., F.T., and A.P. carried out steady-state, photoluminescence and transient optical measurements. The data analysis was performed by G.A. with the supervision of D.C., S.T., and F.Mart.. F.Matt., J.B. and D.D.G. synthesized the materials, fabricated the devices and characterized their electronic properties. All experimental and theoretical findings were collaboratively discussed by the authors. The initial draft of the manuscript was prepared by G.A. and D.C., with all authors contributing to its writing, review, and finalization. S.T., D.C. and A.D.C. are responsible for the founding acquisition.

## Supporting Information

# Effect of FABr Over-Stoichiometry on the Morphology and Optoelectronic Properties of Wide-Bandgap $FAPbBr_3$ Films

G. Ammirati[1], F. Martelli[1], F. Toschi[1], S. Turchini[1], P. O'Keeffe[2], A. Paladini[2], F. Matteocci[3], J. Barichello[1], D. Di Girolamo[3], S. Piccirillo[4], A. Di Carlo[1,3] and D. Catone[1,*].

*[1]Istituto di Struttura della Materia - CNR (ISM-CNR), EuroFEL Support Laboratory (EFSL), Via del Fosso del Cavaliere 100, 00133, Rome, Italy.*

*[2]Istituto di Struttura della Materia - CNR (ISM-CNR), EuroFEL Support Laboratory (EFSL), Monterotondo Scalo 00015, Italy.*

*[3]Centre for Hybrid and Organic Solar Energy (CHOSE), Department of Electronic Engineering, Tor Vergata University of Rome, Via del Politecnico 1, 00133, Rome, Italy.*

*[4]Department of Chemical Sciences and Technologies, Tor Vergata University of Rome Via della Ricerca Scientifica, 00133, Rome, Italy.*

*Corresponding author: daniele.catone@cnr.it.

## Section S.1 - Average Visible Transmittance (AVT)

The average visible transmittance (AVT) of the semitransparent solar cells was determined following the procedure outlined in ISO 9050:2003. The calculation employed the expression:

$$AVT = \frac{\int D(\lambda) \cdot T(\lambda) \cdot V(\lambda) d\lambda}{\int D(\lambda) \cdot V(\lambda) d\lambda}$$

where D($\lambda$) represents the spectral distribution of the incident light (here assumed to be AM 1.5G), T($\lambda$) is the wavelength-dependent transmittance, and V($\lambda$) denotes the photopic response of the human eye.

## Section S.2 - Elliott fit model

The following equation shows the Elliott fitting formula:

$$\alpha_{Elliott}(E) = \alpha_{Cont} + \sum_{n} \alpha_{Ex,n}$$

The first part is the continuous part of the absorption due to the bandgap energy, the second part is the excitonic absorption calculated for $n$ principal quantum numbers. Both contributions are convoluted with broadening functions $\mathcal{N}(energy, central\ energy, broadening\ factor)$ (here a Gaussian). $R_{ex}$ stands for the exciton binding energy, $E_g$ for the bandgap energy, $A_{exc}$ and $A_{cont}$ for the amplitude of the excitonic and continuous absorptions respectively. The broadening parameters are $\sigma_{cont}$ and $\sigma_{exc}$. The coefficient $b$ takes into account the curvature of the continuous part at high energy and is attributed to non-parabolicity of the bands. The equation for absorption we used to fit our data is depicted hereafter, where E and $\epsilon$ are the energy of the absorbed photon:

$$\sum_{n} \alpha_{Ex,n}(E) = A_{exc} \cdot \sum_{n} \left( \frac{1}{E} \cdot \frac{4\pi \sqrt{R_{exc}^3}}{n^3} \right) \otimes \mathcal{N}\left(E, E_g - \frac{R_{exc}}{n^2}, \sigma_{exc}\right)$$

$$\alpha_{Cont}(E) = A_{cont} \cdot \int_{E_g}^{\infty} \left( \frac{1}{\epsilon} \cdot \frac{\pi \exp(\pi x)}{\sinh(\pi x)} \cdot \sqrt{\epsilon - E_g} \cdot \frac{1}{1 - b\left(\epsilon - E_g\right)} \right) \otimes \mathcal{N}(E - \epsilon, 0, \sigma_{cont}) d\epsilon$$

Where

$$x = \sqrt{\frac{R_{exc}}{\epsilon - E_g}}$$

## Section S.3 - Fast Transient Absorption Spectroscopy

This interpretation is further supported by FTAS measurements performed with a pump energy of 2.48 eV and a pump fluence of 66 μJ/cm². **Figure S1a** reports the normalized transient absorption spectra for Control, FABr +5% and FABr +10%, acquired with a pump energy of 2.48 eV and a pump fluence of 66 μJ/cm². The transient spectra exhibit a clear photobleaching signal at energies closely matching the optical onsets observed in the steady-state absorbance spectra shown in **Figure 1**. The spectral profiles of the transient signals are characteristic of excitonic transitions, in good agreement with the PL results discussed in the main text. In addition, the decay dynamics of all materials exhibit comparable decay trends with lifetimes on the order of several hundreds of picoseconds or nanoseconds, as shown in **Figure S1b**. This behavior is due to the fact that the FTAS is not so sensible to the defectivity of the material but more to the exciton recombination, reinforcing the results obtained with PL.

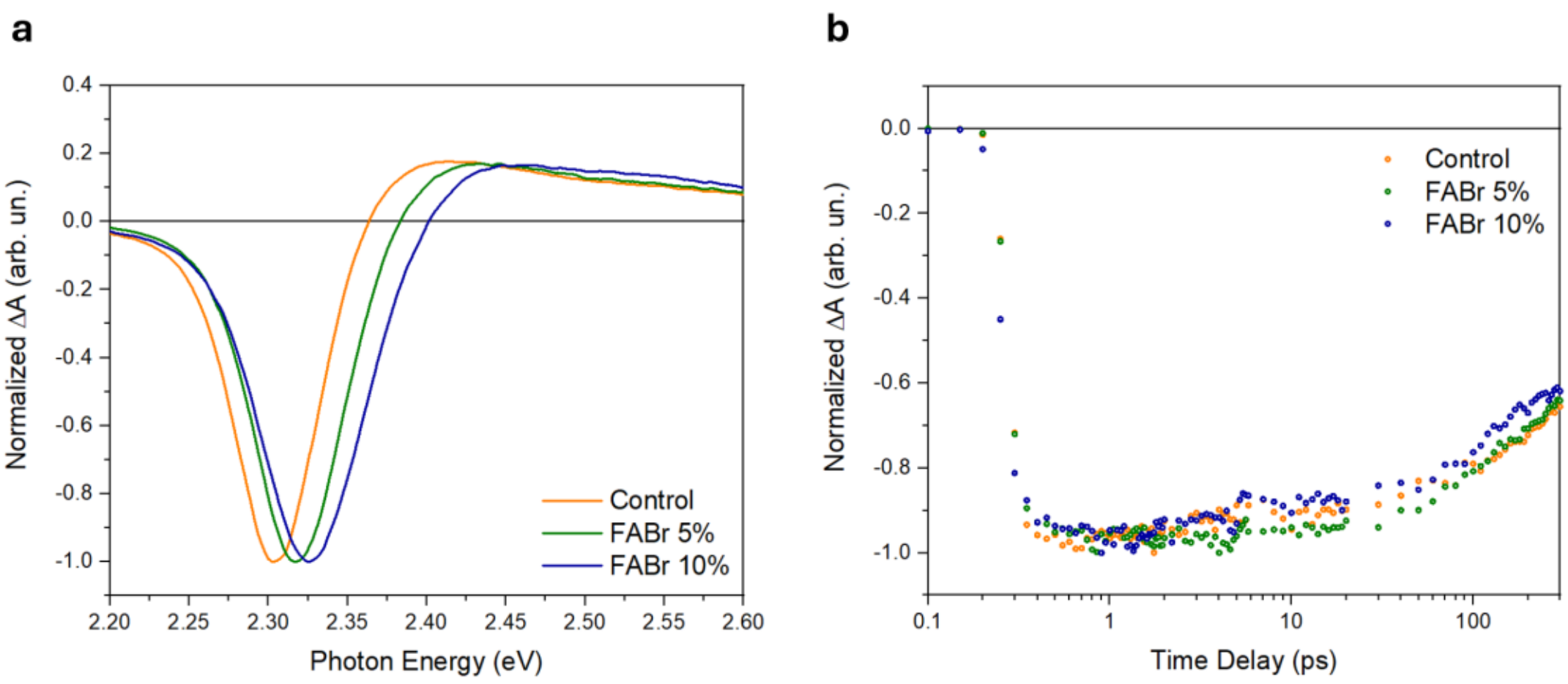

**Figure S1:** (a) Transient spectra and (b) decay dynamics acquired with a pump energy of 2.48 eV and a pump fluence of 66 μJ/cm² for Control, FABr +5% and FABr +10%.

## Section S.4 - Temperature-dependent PL

**Figure S2a** and **S2b** show the PL spectra acquired at selected temperatures ranging from 12 to 330 K for the Control and FABr +10% samples. The dotted colored lines are included as guides for the eye to highlight the shifts of the PL peak maxima across the measured temperature range. Both the FABr +10% and Control samples exhibit an overall blueshift of the PL emission with increasing temperature, a characteristic behavior widely documented for halide perovskites, typically originated from the reverse temperature-dependent ordering of the electronic band edge.[1]

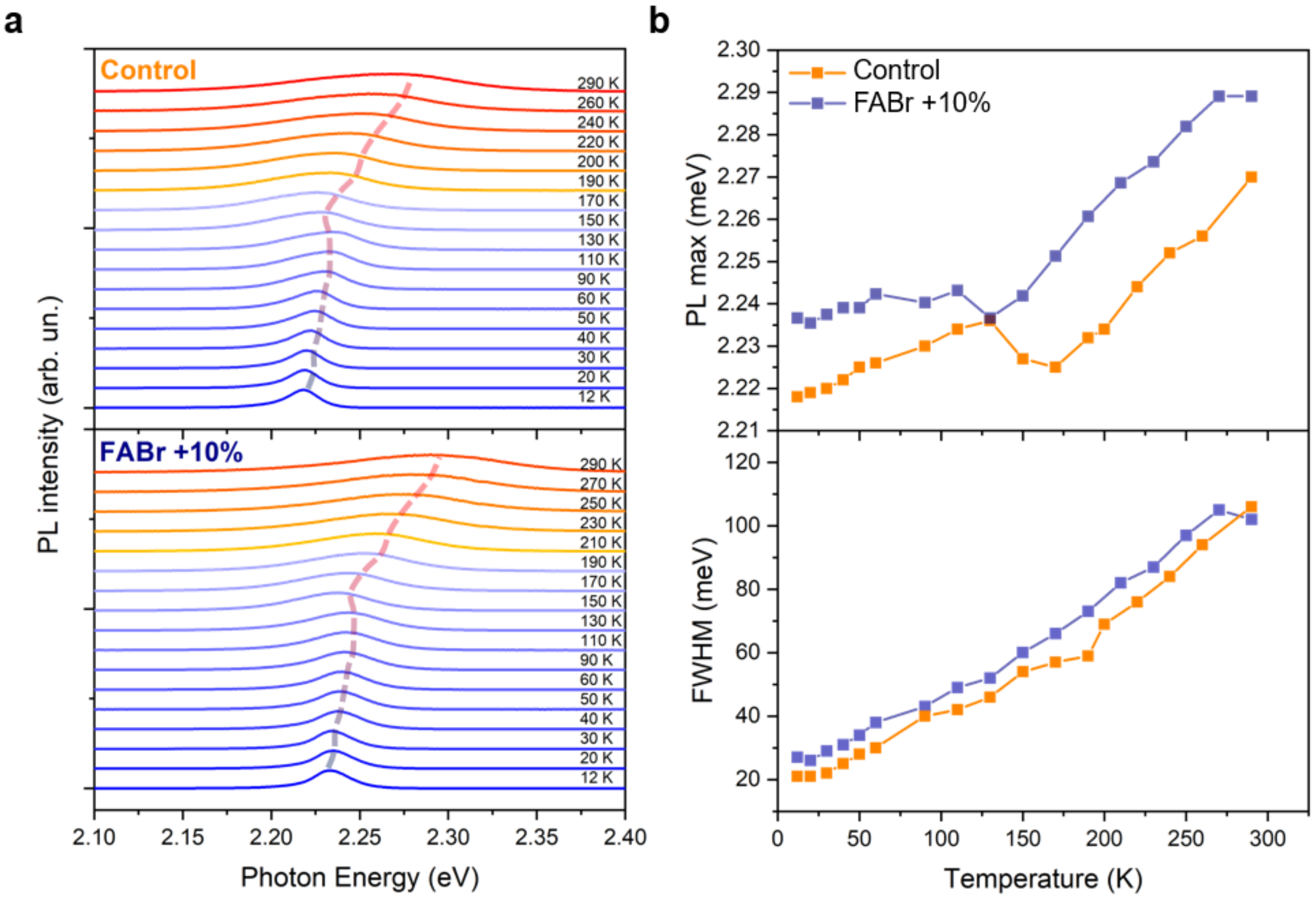

**Figure S2:** (a) PL spectra obtained at selected temperatures between 12 and 300 K at an irradiance of 100 mW/cm² for (a) the Control and (b) FABr +10% samples; the dotted lines are included as guides for the eye to highlight the shifts of the PL maxima across the measured temperature range, which indicates a blueshift as the temperature increases.

**Figure S2c** and **S2d** show the temperature dependence of the PL peak energy and the full width at half maximum (FWHM) for the Control and FABr +10% samples. Both trends confirm the overall blueshift of the emission peak with increasing temperature, as previously discussed. Notably, a clear discontinuity is observed at about 130 K, consistent with structural phase transitions that are characteristic of formamidinium-based lead halide perovskites, namely the orthorhombic-to-tetragonal phase transition.[2–4]

The increase in the FWHM of the PL spectra with rising temperature is an expected behavior and reflects the thermally induced broadening of emission lines.[3] Notably, the FABr +10% sample exhibits a systematically larger FWHM compared to the Control over the entire temperature range. This difference can be ascribed to increased compositional disorder, arising from the random positioning of the FA excess in the lattice. The excess FABr introduces a broader distribution of local environments in the perovskite lattice, leading to inhomogeneous broadening of the PL signal. Interestingly, while the introduction of FABr excess reduces defect-related recombination, as evidenced by the absence of low-energy PL features and the transition to excitonic radiative behavior, it also enhances the degree of structural disorder. This dual effect suggests that FABr excess plays a complex role in tuning the optoelectronic properties of the material.

## Section S.5 – XRD

**Figure S3** shows the XRD patterns collected for Control, FABr +5% and FABr +10%.

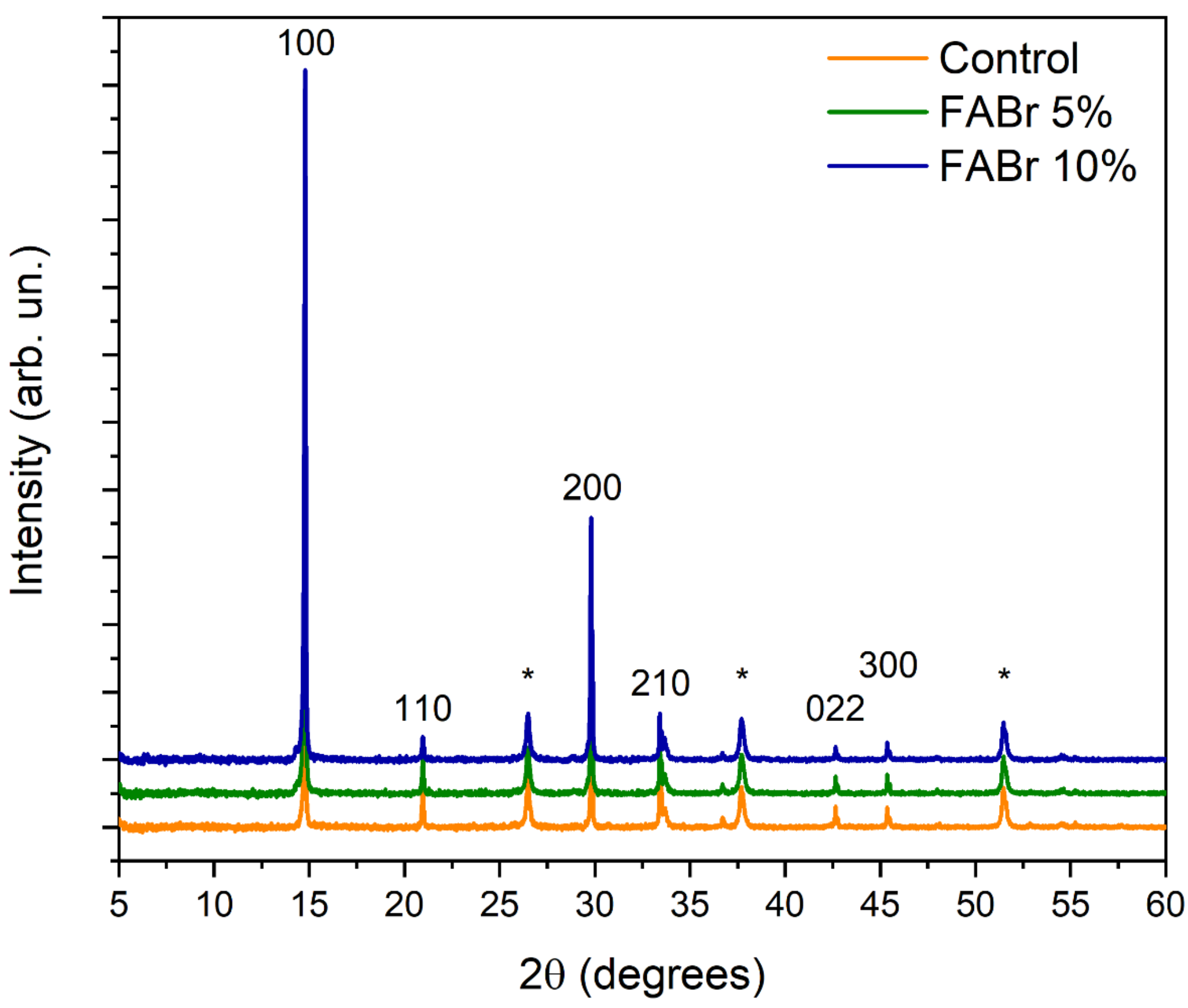

**Figure S3:** XRD patterns of Control, FABr +5% and FABr +10%. All samples exhibit the characteristic reflections of the $\alpha$-$FAPbBr_3$ perovskite phase. The main diffraction peaks are indexed with the corresponding Miller indices. Peaks marked with an asterisk (*) originate from the FTO substrate.

## Section S.6 - PV Performance

Current density–voltage (J–V) characteristics were recorded in ambient conditions using a calibrated ABET Sun 2000 solar simulator (Class A), operating under AM 1.5 illumination at an intensity of 100 mW $cm^{-2}$. The system was calibrated with a certified silicon reference device (RERA Solutions RR-1002). Measurements were carried out by sweeping the bias in both forward and reverse directions, employing a scan rate of 200 mV $s^{-1}$ and a voltage increment of 20 mV. A multichannel four-probe source-meter (Arkeo-Ariadne, Cicci Research srl) was used to conduct the electrical characterization.

External Quantum Efficiency (EQE) spectra were obtained with an independent commercial measurement platform (Arkeo-Ariadne, Cicci Research srl) equipped with a 300 W xenon light source. The system enables spectral measurements in the 300–1100 nm range with a wavelength resolution of 2 nm.